\RequirePackage[2020-02-02]{latexrelease}
\documentclass[aps,pra,showpacs,superscriptaddress,10pt,twocolumn]{revtex4}
\usepackage{amsmath}
\usepackage{amsfonts}
\usepackage{graphicx}
\usepackage{float}
\usepackage{color}
\usepackage{doi}
 \usepackage{bm}
 \usepackage{array}
 \usepackage[T1]{fontenc}
 \usepackage{bigints}
 \usepackage{url}
\usepackage{natbib}

\begin{document}

\title{Squeezing properties of degenerate high-order hyper-Raman lines emitted by a two-level system}

\author{\'{A}kos Gombk\"{o}t\H{o}}
\affiliation{University of Pannonia, Zalaegerszeg Campus,	Zalaegerszeg, H-8900 Hungary}
\email{gombkoto.akos@zek.uni-pannon.hu}
\affiliation{Wigner Research Centre for Physics, Konkoly-Thege	M. \'ut 29-33, H-1121 Budapest, Hungary}

\begin{abstract}
Quantized descriptions of nonlinear-optical processes can be relevant from the perspective of developing novel nonclassical sources of light. As a special case, it is useful to characterize light emitted by classically driven systems, since the excitations in many practical cases are laser beams. As a material model, we choose a two-level system.
In an earlier work, we investigated photon statistics and intermodal cross-correlations Phys. Rev. A \textbf{104}(3):033703, (2021), and noted that squeezing is primarily present in specific sets of parameters, corresponding to the degeneracy of hyper-Raman lines. Here we focus on this specific set of parameters, presenting details of the squeezing. 
%By generalizing the material system to an ensemble of 2-level atoms, we conclude that the many-atom effects can enhance the resulting squeezing. 
\end{abstract}
%\pacs{}
\maketitle 

\section{Introduction} \label{intro}

The concept of squeezed light is a historically important idea. Not only fundamental but also applied science has benefitted a lot  from  using squeezed light, being  a valuable tool of metrology and quantum information processing \cite{XK89,PK92,BRL19,GB89,NNW03,RSP17}.

A light field is analyzed with respect to many different modulation frequencies, and the result constitutes the squeezing spectrum \cite{BIS07}, where both the degree and the bandwidth of squeezing are important for potential applications \cite{YYR17}. 
While experiments with pulsed squeezed light reach a bandwidth up to tens of THz  \cite{IC09}, it may be of practical interest to develop sources of squeezed light characterized by ever greater bandwidth.

\bigskip

From another direction, in recent years there has been an increasing interest in quantum optical descriptions of highly nonlinear optical phenomena, particularly toward exploring nonclassical states of the electromagnetic field induced by high-field light-matter interaction \cite{photonics8070269,LMCMP21,TNKIGIT17,TKDF19}.
A groundbreaking work \cite{AG19} have shown qualitatively that nonclassical properties can be expected in the harmonic radiation even for rather general material systems and excitations. 
A characterization of the harmonics was given for the case of monochromatic excitation and two-level system \cite{GFV21Q} have shown nonclassical features including squeezing being present in the scattered modes, particularly in the hyper-Raman lines for the case of a given choice of parameters, when these lines are degenerate. 

\bigskip

We note in advance, that the squeezing in the optical lines of a single two-level atom is going to be negligible, especially compared to state-of-the-art sources of squeezed light. However, there are a few points that are worth raising:

i) As shown in \cite{AG19}, for a manifold of identical 2-level systems, the squeezing present in the scattered radiation can be significant, and can be increased further with the number characterizing the size of the manifold.

ii) While the squeezing in the optical lines of a single 2-level atom is weak, the number of lines (and the spectral width of the squeezing spectrum) can be increased, in principle arbitrarily.

iii) High-intensity quantum optical devices can be extremely flexible in their application (one-photon states, multimode entanglement, squeezing in the harmonics, Schrödinger-cat states in the excitation) but it is important to identify the limits of each specific application so that long-term expectations can be realistically evaluated.

\section{Model}

The fundamental, and necessarily very simplified model of the material system in quantum optics is the two-level atom, which may be acceptably used to describe harmonic generation in semiconductor heterostructures \cite{H94,HG22}, granted that the wavefunction spams essentially two isolated bound states, with negligible effect on the dynamics by additional components. On the other hand, the simplicity of two-level systems helps forming qualitatively correct predictions and offers insight into the dynamics as well. 

In our model, we will assume that the external driving can be treated classically and that the dipole approximation is correct. 
We will consider the Hamiltonian to be independent of spatial coordinates, and neglect the effects due to the propagation of the field \cite{GC16,photonics8070263,G20}.
Let us consider the following terms:
\begin{equation}
H_{a}=\hbar\frac{\omega_{0}}{2}\sigma_z, 
\end{equation}
\begin{eqnarray}
	H_{m}=\sum_n \hbar \omega_n a_n^{\dagger}a_n,
~~~~
H_{am}=\sum_n \hbar \frac{\Omega_n}{2}\sigma_x (a_n+a_n^{\dagger}),
\label{Hint}
\end{eqnarray}
which corresponds to the two-level atom, the quantized electromagnetic modes, and the quantized dipole interaction, respectively.
Since the strong external exciting pulse can be described classically, we use: 
\begin{equation}
H_{ex}(t)= -DE(t)=-d\sigma_x E(t)=-\hbar\frac{\Omega(t)}{2} \sigma_x.
\end{equation}
We note that $\Omega_n=2d\sqrt{\frac{\hbar \omega_n}{\epsilon_0 V}}$, where $V$ is the quantization volume.
Let us denote the eigenstates of the atomic Hamiltonian by $|e\rangle$ and $|g\rangle$, 
i.e., $H_a|e\rangle=\hbar\omega_0/2 |e\rangle,$ $H_a|g\rangle=-\hbar\omega_0/2 |g\rangle.$ 
\\
In the following we will investigate the system:
\begin{equation}
H(t)=H_{a}+H_{m}+H_{am}+H_{ex}(t)
\label{Ham},
\end{equation}
with the initial state being $|\Psi_0 \rangle = |g\rangle\otimes\prod_n|0\rangle$, which is approximately the ground state of the system. Throughout this paper, we used $\Omega_n/\sqrt{\omega_n}=0.005$ parameter.

\subsection{Measure of squeezing}\label{corrfunc}

We use the notations below:
\begin{eqnarray*}
	X_n=\frac{a^\dagger_n + a_n}{2} 
	~~~,~~~
	Y_n=i\frac{ a^\dagger_n - a_n}{2} 
	\\
	X_{2_n}=\frac{a^{\dagger 2}_n + a^2_n}{2} 
	~~~,~~~
	Y_{2_n}=i\frac{ a^{\dagger 2}_n - a^2_n}{2} 
\end{eqnarray*}
Squeezed  states  are  associated  with canonical observables, in quantum optics typically electric field strength at a given $\phi$ phase. The corresponding dimensionless 
$X^\phi_n$ and $Y^\phi_n$ operators are defined as: 
\begin{eqnarray*}
	X^\phi_n = \frac{a_n + a^\dagger_n}{2}\cos\phi +  i\frac{a^\dagger_n-a_n}{2} \sin\phi,
	\\
	Y^\phi_n = -\frac{a_n + a^\dagger_n}{2}\sin\phi +  i\frac{a^\dagger_n-a_n}{2} \cos\phi,
\end{eqnarray*}
and for the sake of completeness, we write the quadratic variances as:
\begin{eqnarray*}
	\langle (\Delta X^\phi)^2 \rangle 
	= \frac{1}{4}\bigg( 1 + 2\langle N \rangle + 2\langle X_2\rangle - 4\langle X\rangle^2 \bigg)\cos^2\phi
	\\ +
	\frac{1}{4}\bigg( 1 + 2\langle N \rangle - 2\langle X_2\rangle - 4\langle Y\rangle^2 \bigg)\sin^2\phi
	\\ +
	\bigg( \langle Y_2\rangle - 2\langle Y\rangle \langle X\rangle \bigg)\cos\phi \sin\phi
	, \end{eqnarray*}
\begin{eqnarray*}
	\langle (\Delta Y^\phi)^2 \rangle 
	= \frac{1}{4}\bigg( 1 + 2\langle N \rangle + 2\langle X_2\rangle - 4\langle X\rangle^2 \bigg)\sin^2\phi
	\\ +
	\frac{1}{4}\bigg( 1 + 2\langle N \rangle - 2\langle X_2\rangle - 4\langle Y\rangle^2 \bigg)\cos^2\phi
	\\ -
	\bigg( \langle Y_2\rangle - 2\langle Y\rangle \langle X\rangle \bigg)\cos\phi \sin\phi .
\end{eqnarray*}

Light is considered squeezed if there exists a mode $n$ and phase $\phi$ \cite{WM94,CMW85} such that $\Delta X^{\phi}_n<\frac{1}{2}$ .
The minimal variance (and its associated phase) can be calculated through  the smaller eigenvalue (and associated eigenvector) of the noise-ellipse matrix:
\begin{equation}
	\left[  
	\begin{matrix}
		\langle (\Delta X)^2 \rangle  & \hspace{0.5cm} \frac{1}{2}\langle\{\Delta X, \Delta Y\}\rangle
		\\
		\frac{1}{2}\langle\{\Delta X, \Delta Y\}\rangle & \hspace{0.5cm} \langle (\Delta Y)^2 \rangle
	\end{matrix}\right]
\end{equation}
The eigenvalues \cite{PJ91} and eigenvectors \cite{LPP88}, expressed with our notations are:
\begin{eqnarray*}
	\lambda_{\pm}=
	\frac{1}{4}\bigg[ 
	\langle \{ \Delta a, \Delta a^\dagger \} \rangle
	\pm 2 |\langle (\Delta a)^2 \rangle|
	\bigg]
	\nonumber \\
	= \frac{1}{4}\bigg[ 
	1 \!+ \!2\big(\langle N \rangle \!-\! \langle X \rangle^2 \!-\! \langle Y \rangle^2 \big)
	\!\pm 2 | \langle X_2 \! + \! i Y_2 \rangle \!-\! \langle X \!+\! i Y \rangle^2 |
	\bigg] ,
\end{eqnarray*}
and $\overline{u}_+$, $\overline{u}_-$
with the components fulfilling:
\begin{eqnarray*}
	(u_\pm)^2_1 \!=\! 
	\frac{\big( \lambda^{\pm}  \!- \! \langle 2 X_2 \!+\! 2N \!+\! 1  \rangle 
		\!-\! 4\langle Y \rangle^2 \big)^2 }
	{\big( \lambda^{\pm}  \!- \! \langle 2 X_2 \!+\! 2N \!+\! 1  \rangle 
		\!-\! 4\langle Y \rangle^2 \big)^2 \!+\! \langle 2Y_2 \! -\! 4 \langle X \rangle \langle Y \rangle \rangle^2}
	\nonumber \\
	(u_\pm)^2_2 \!=\! 
	\frac{\langle 2Y_2 \! -\! 4 \langle X \rangle \langle Y \rangle \rangle^2}
	{\big( \lambda_{\pm}  \!- \! \langle 2 X_2 \!+\! 2N \!+\! 1  \rangle 
		\!-\! 4\langle Y \rangle^2 \big)^2 \!+\! \langle 2Y_2 \! -\! 4 \langle X \rangle \langle Y \rangle \rangle^2}.
\end{eqnarray*}
To quantify the squeezing of a given mode, we will also use the notation
\begin{equation}\vspace*{-0.5cm}
	S\equiv \log_{10}(4\lambda_{-}).
\end{equation}

\section{Dynamics of the degenerate hyper-Raman lines}

\begin{figure}[h]
	\centering
	\includegraphics[width=0.85\linewidth]{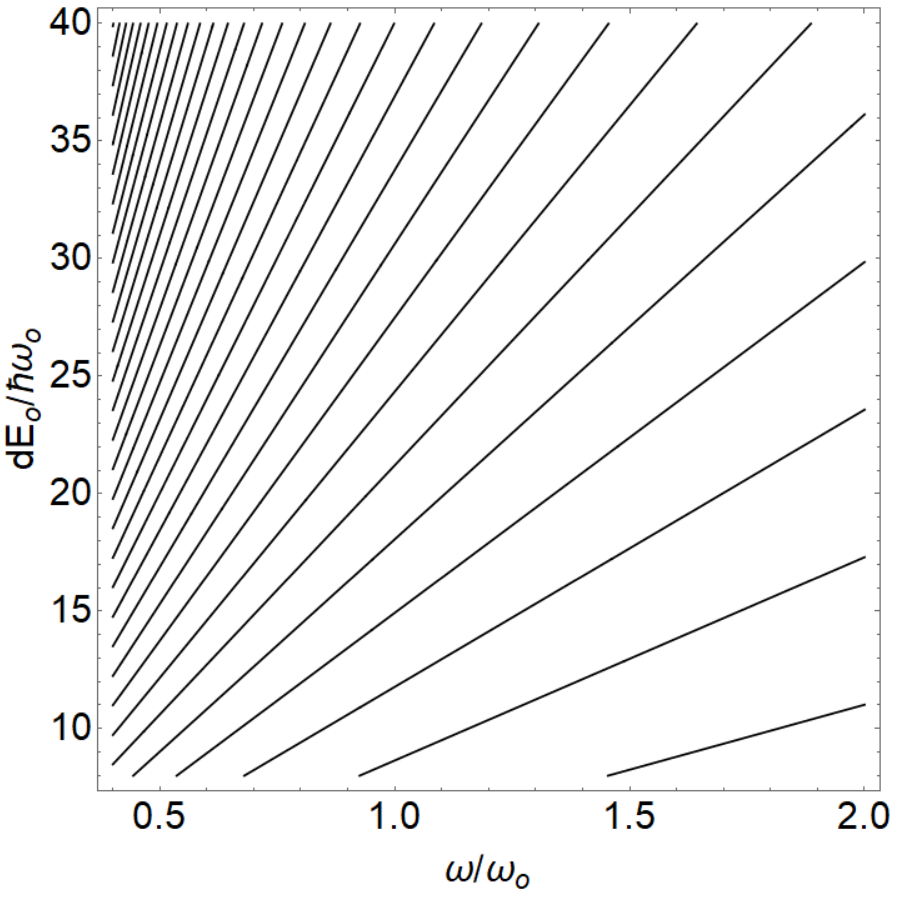}
	\caption{Black lines show the parameters corresponding to the special cases when the hyper-Raman lines correspond to even harmonic frequencies. The axis corresponds to the dimensionless amplitude and frequency of the excitation. (based on the calculations in the appendix of \cite{GFV21Q})}
	\label{fig:omegazeropar}
\end{figure}

Let us remind the reader that the harmonic spectra of monochromatically driven two-level atoms contain odd-harmonic lines and Mollow-sidebands/hyper-Raman lines \cite{MWENO05}, which we have also referred to as even-order harmonics in \cite{GC16,GFV21Q} due to their proximity to the even multiples of the base harmonic. 
\\In this work, we focus on the special case when the frequency of these lines exactly coincides with the even multiples of the driving frequency. An illustrative plot of the relevant parameters is shown in Fig.(\ref{fig:omegazeropar}).

\bigskip

\begin{figure}[h!]
	\centering
	\includegraphics[width=1.0\linewidth]{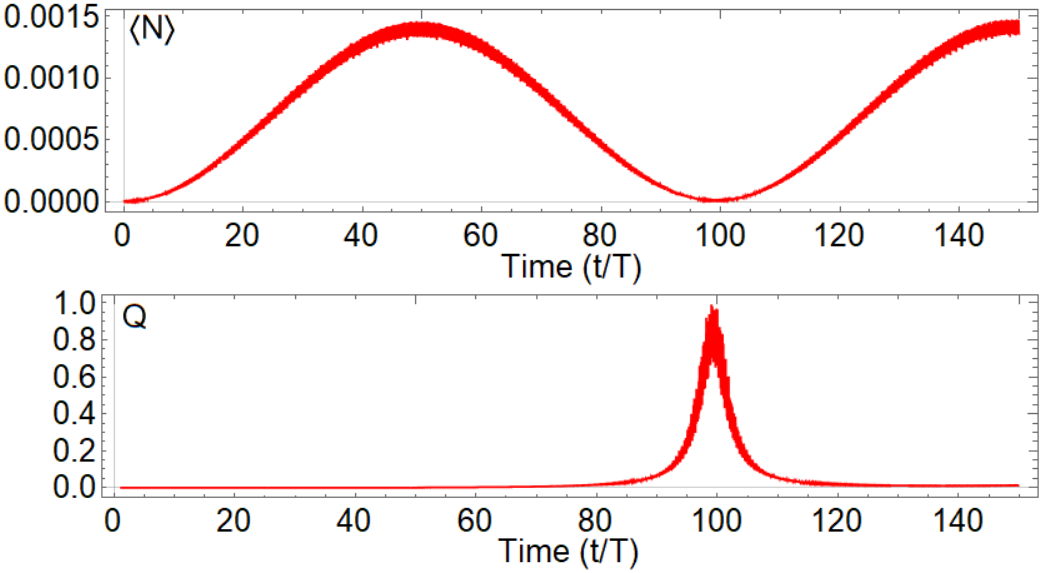}
	\caption{Illustrative plot of the dynamical evolution of photon number expectation value (top panel) and Mandel-parameter (bottom panel) of the 8th harmonic under monochromatic excitation.}
	\label{fig:evenharmn}
\end{figure}

It is worth pointing out, that in our experience --under monochromatic excitation-- practically all optical lines are characterized by photon numbers monotonically increasing in time, except for the case of degenerate hyper-Raman lines (that is, when the frequency of the two spectrally nearest sidelines coincide). The anomalous time-evolution of these modes is such, that the population fluctuate in time, see Fig.(\ref{fig:evenharmn}).

\pagebreak

\vspace*{-1.5cm}

The dynamical fluctuation has a more-or-less well-defined "period" although of course it is not periodic in the strict sense. The initial vacuum state gains noticeable modification during the period. We can make the following statement: For all degenerate hyper-Raman lines, both the qualitative features and the approximate period are the same. 
This implies that after the isolation of the Raman-lines from the odd-order harmonics, (which may be possible due to higher spatial divergence \cite{BEBDPLMM19}) ultrashort squeezed pulses may be generated.

The dependence of the period on the system's parameters on the other hand is highly nontrivial. An analytical calculation regarding that can be seen in the Appendix.

\begin{figure}[h!]
	\centering
	\includegraphics[width=0.99\linewidth]{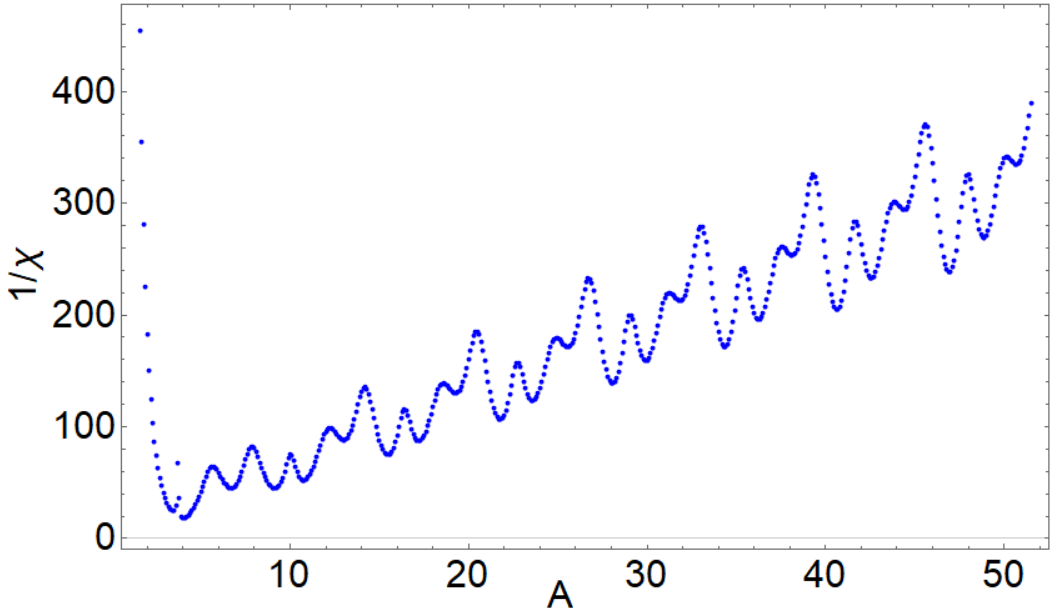}
	\caption{Illustrative plot of the approximate period measured in units of optical periods of the excitation, for resonant excitation, as a function of the (dimensionless) excitation amplitude $A$. ($1/\chi$ as per the nomenclature of the appendix.)}
	\label{fig:period}
\end{figure}

\subsection{Quadrature-squeezing time-evolution and spectra}\label{dynequs}
\begin{figure}[h!]
	\centering
	\includegraphics[width=0.99\linewidth]{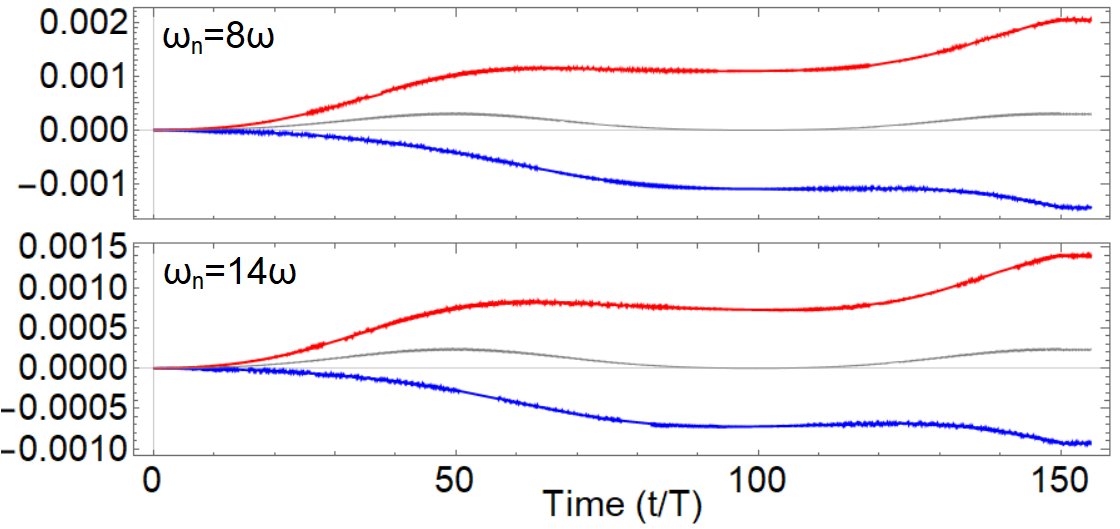}
	\caption{Time-evolution of $\lambda_{-}\!-\!\frac{1}{4}$ (blue), $\lambda_{+}\!-\!\frac{1}{4}$ (red) and $\sqrt{\lambda_{+}\lambda_{-}}\!-\!\frac{1}{4}$ (gray) values characterizing the 8th and 14th harmonic, respectively.}
	\label{fig:squeeze5}
\end{figure}

Below, we considered an excitation that is monochromatic for a given number of optical cycles, before the amplitude cuts off. Time-evolution of relevant quantities for two given (8th and 14th) harmonic modes can be seen in Fig.(\ref{fig:squeeze5}).
\\
Numerical calculations indicate that the squeezing becomes more significant for longer interaction times and larger coupling strengths. At the same time, it is apparent that most of the time $\sqrt{\lambda_{+}\lambda_{-}}>\frac{1}{4}$ is fulfilled, meaning that the quantum states characterizing even harmonic modes, while squeezed, are not intelligent states. Hence, they cannot be correctly described by squeezed coherent states.

On the other hand, at the interaction time when $\langle N \rangle$ reaches a local minimum and $Q\approx 1$, the quantum state can be considered to be practically a squeezed vacuum state.

\pagebreak

Under an excitation with rectangular carrier function, with a cutoff at $t\approx 50T$, the quantities follow an oscillatory dynamics around a well-defined value. We calculated this through averaging, and plotted the asymptotical value of $\lambda_{-}-\frac{1}{4}$ as a function of frequency in Fig.(\ref{fig:squeezingspectra}).
We note that the squeezing spectrum is qualitatively similar, (albeit less structured, and more complicated) in the case of pulsed excitations.
\begin{widetext}
\begin{minipage}{\linewidth}\vspace{-0.4cm}
\begin{figure}[H]
	\centering
	\includegraphics[width=0.90\linewidth]{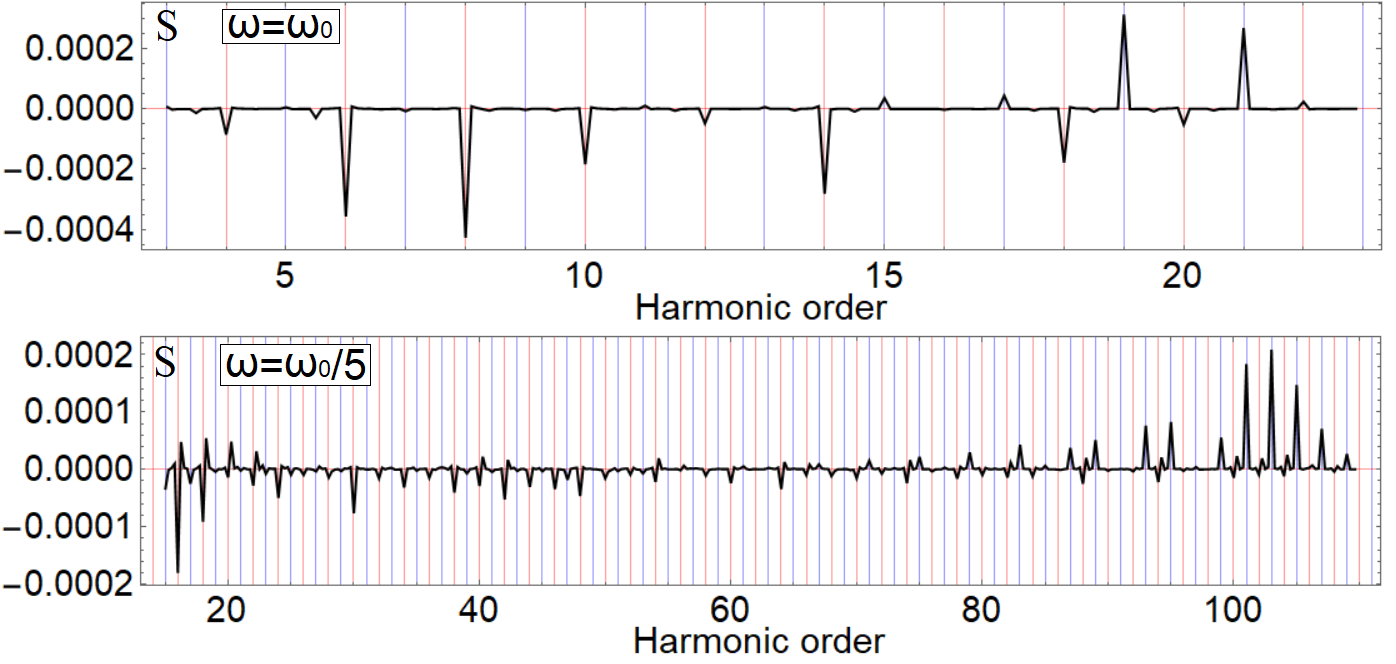}
	\caption{Squeezing spectra with resonant and red-detuned excitation. Colored vertical lines show the position of odd-order (blue), and even-order (red) harmonics. One can observe that squeezing is present primarily in even harmonics.}
	\label{fig:squeezingspectra}
\end{figure}
\end{minipage}
\end{widetext}

\subsection{Characterization of squeezing}

From a practical point of view, not just the presence of the squeezed quantum states is important, but whether they correspond to known special cases, such as amplitude- or phase squeezed states. 
In order to gain insight we define the quadrature angle as:
\begin{equation*}
\varphi \equiv atan2 \big( \langle Y\rangle, \langle X \rangle \big),
%\chi\equiv \text{atan2}\big( (u_-)_2, (u_-)_1 \big),
\end{equation*}
and compare the time-evolutions of $\langle(\Delta X^\varphi)^2\rangle$ and $\langle(\Delta Y^\varphi)^2\rangle$. We note that --since the photon quadrature-averages do not reach significant values-- these quantities can only qualitatively be interpreted as amplitude and phase variances.

In our experiences, while the time-evolution of $\langle(\Delta X^\varphi)^2\rangle-\frac{1}{4}$ and $\langle(\Delta Y^\varphi)^2\rangle-\frac{1}{4}$
is complicated, it is qualitatively very similar for all degenerate even harmonics to that seen in Fig.(\ref{fig:evenphases}).
\\
A qualitative statement that can be made is that the degenerate hyper-Raman photons can be approximately either phase-squeezed, amplitude-squeezed, or squeezed vacuum, depending on the interaction time. 

\begin{widetext}
	\begin{minipage}{\linewidth}	
		\begin{figure}[H]
			\centering
			\includegraphics[width=0.85\linewidth]{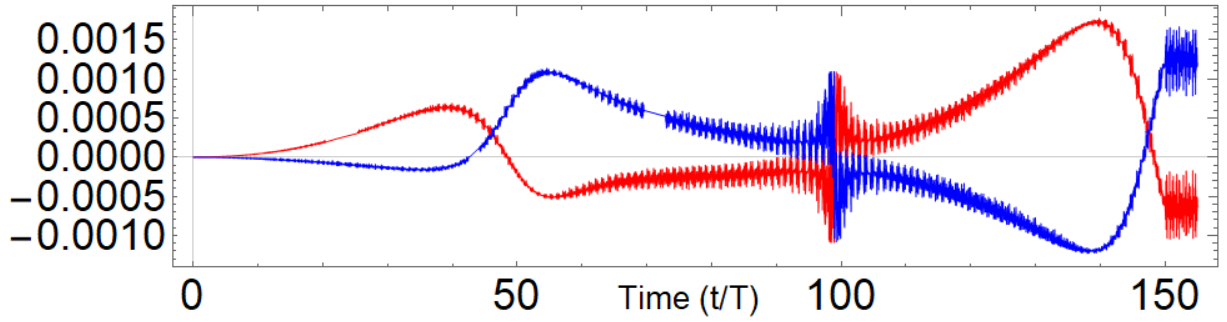}
			\caption{Time-evolution of $\langle(\Delta X^\varphi)^2\rangle-\frac{1}{4}$ (red) and $\langle(\Delta Y^\varphi)^2\rangle-\frac{1}{4}$ (blue) of the 8th harmonic. The monochromatic excitation has a cutoff after 150 optical cycles.}
			\label{fig:evenphases}
		\end{figure}
	\end{minipage}
\end{widetext}

For short interaction times, we can expect the degenerate hyper-Raman lines to be approximately phase-squeezed. Since the dynamics coincide for all Raman lines, we can expect that the generation of wideband phase-squeezed pulses should be possible.

\section{Conclusions} \label{conclusion}

We identified the parameters for which the hyper-Raman lines of the HHG spectra emitted by a two-level system are degenerate. Our calculations suggest that in this special case, the corresponding degenerate modes are in a significantly more squeezed state than for other parameters. Generally speaking, we can conclude that the squeezing increases with interaction time, and depending on its exact value, the squeezed states can be characterized as either amplitude-squeezed, phase-squeezed, or squeezed vacuum. 
The type of squeezing coincides for all degenerate hyper-Raman lines, and is independent of the modal frequency and coupling strength.
%We wish to point out that the calculations in this paper can be generalized to multi-level systems in a straightforward manner.

\begin{acknowledgements}
	We are grateful for the support of the National Research, Development and Innovation Office of Hungary (Project No. K124351 and TKP2021-NVA-04).
\end{acknowledgements}

\appendix

\section{Analytical results}\label{Analyticalkorrelacio}
%\subsection{Approximation of photon statistics}

Since the photon statistics of the harmonics differ little from the vacuum state, a perturbative treatment of the dynamics in independent-mode approximation seems to be reasonable, at least at a qualitative level. With the same transformations and notations as presented in \cite{GFV21Q}, the wavefunction can be written in the following form:
%\begin{widetext}
\begin{eqnarray*}\label{wavefunc}
|\Psi\rangle =
e^{i\frac{A\xi}{2\omega} \sin(\omega t + \phi_0 )\sigma_x} e^{\frac{i}{2}(\omega t+\phi_0)\sigma_z} e^{-\frac{i}{\hbar}t\epsilon_+}
\sum^{\infty}_{j=0} b^e_j | \tilde{e} \rangle|j \rangle e^{-ij \omega_n t}
\nonumber \\+
e^{i\frac{A\xi}{2\omega} \sin(\omega t + \phi_0)\sigma_x}
e^{\frac{i}{2}(\omega t+\phi_0)\sigma_z} e^{-\frac{i}{\hbar}t\epsilon_-}
\sum^{\infty}_{j=0} b^g_j | \tilde{g} \rangle|j \rangle e^{-ij \omega_n t} ,
\end{eqnarray*}
%\end{widetext}
with the dynamical equations governing the $b^{g}_j, b^{e}_j$ coefficients turn out to be:
\begin{eqnarray*}
i \dot{b}^e_j(t) = \langle \tilde{e} | W(t) | \tilde{e} \rangle b^e_j(t)
+ \langle \tilde{e} | W(t) | \tilde{g} \rangle 
e^{i\frac{\epsilon_+ - \epsilon_-}{\hbar}t} ~ b^g_j(t) 
\\+ \frac{\Omega}{2} 
e^{i(\delta\omega t - \phi_0) } 
\sum_k \langle j|  a + a^\dagger | k\rangle e^{-i\omega_n(k-j)t}  ~b^g_k(t)
\\- \Omega \cos\theta 
e^{i\frac{\epsilon_+-\epsilon_-}{\hbar}t } \cos(\omega t+\phi_0)
\\\times
\sum_k \langle j|  a + a^\dagger | k\rangle e^{-i\omega_n(k-j)t}  ~b^g_k(t)
\\+ \frac{\Omega}{2} \sin(2\theta) \cos(\omega t+\phi_0)
\\\times
\sum_k \langle j|  a + a^\dagger | k\rangle e^{-i\omega_n(k-j)t}  ~b^e_k(t)
,
\end{eqnarray*}
\begin{align*}
i \dot{b}^g_j(t) = \langle \tilde{g} | W(t) | \tilde{g} \rangle b^g_j(t)
+ \langle \tilde{g} | W(t) | \tilde{e} \rangle e^{-i\frac{\epsilon_+ - \epsilon_-}{\hbar}t} ~b^e_j(t) 
\\+ \frac{\Omega}{2} 
e^{-i(\delta\omega t + \phi_0)} 
\sum_k \langle j|  a + a^\dagger | k\rangle e^{-i\omega_n(k-j)t}  ~b^e_k(t) 
\\- \Omega \cos\theta 
e^{i\frac{\epsilon_--\epsilon_+}{\hbar}t } \cos(\omega t+\phi_0)
\\\times
\sum_k \langle j|  a + a^\dagger | k\rangle e^{-i\omega_n(k-j)t}  ~b^e_k(t)
\\- \frac{\Omega}{2} \sin(2\theta) \cos(\omega t+\phi_0)
\\\times
\sum_k \langle j|  a + a^\dagger | k\rangle e^{-i\omega_n(k-j)t}  ~b^g_k(t).
\end{align*}
These general dynamical equations can be slightly simplified in the special case explored in this work, as below. 
%We will treat two cases (odd-harmonic modes $\omega_n=(2k+1)\omega$ and hyper-Raman lines $\omega_n=2k \omega$ separately. 
We note here that resonances in the first-order of perturbation-calculations are present for $\omega_n=0$, $\omega_n=\omega$, $\omega_n=2\omega$ modes, however, we will exclude these in favour of the higher harmonic modes.

Here we only use analytical reasoning to gain an understanding of the behaviour visible on Fig.($\ref{fig:evenharmn}$) and Fig.($\ref{fig:squeeze5}$), that is, that degenerate hyper-Raman lines have an approximately periodic time-evolution, at the end of which its quantum state corresponds to that of a squeezed vacuum state. Specifically, we shall calculate analytically this period, which turns out to be independent of the harmonic order and coupling strength.

\begin{widetext}
	\begin{align}
		i \dot{b}^e_j(t) = \langle \tilde{e} | W(t) | \tilde{e} \rangle b^e_j(t)
		+ \langle \tilde{e} | W(t) | \tilde{g} \rangle 
		e^{i \omega t} ~ b^g_j(t) 
		+ \frac{\Omega}{2} e^{-i\phi_0 } 
		\sum_k \langle j|  a + a^\dagger | k\rangle e^{-i\omega_n(k-j)t}  ~b^g_k(t)
	\nonumber	\\- \Omega \cos\theta 
		e^{i\omega t} \cos(\omega t+\phi_0)
		\sum_k \langle j|  a + a^\dagger | k\rangle e^{-i\omega_n(k-j)t}  ~b^g_k(t) + \frac{\Omega}{2} \sin(2\theta) \cos(\omega t+\phi_0)
		\sum_k \langle j|  a + a^\dagger | k\rangle e^{-i\omega_n(k-j)t}  ~b^e_k(t) ,
	\end{align}
	\begin{align}
		i \dot{b}^g_j(t) = \langle \tilde{g} | W(t) | \tilde{g} \rangle b^g_j(t)
		+ \langle \tilde{g} | W(t) | \tilde{e} \rangle e^{-i\omega t} ~b^e_j(t) 
		+ \frac{\Omega}{2} 
		e^{-i\phi_0} 
		\sum_k \langle j|  a + a^\dagger | k\rangle e^{-i\omega_n(k-j)t}  ~b^e_k(t) 
	\nonumber	\\- \Omega \cos\theta 
		e^{-i\omega t} \cos(\omega t+\phi_0)
		\sum_k \langle j|  a + a^\dagger | k\rangle e^{-i\omega_n(k-j)t}  ~b^e_k(t)
		- \frac{\Omega}{2} \sin(2\theta) \cos(\omega t+\phi_0)
		\sum_k \langle j|  a + a^\dagger | k\rangle e^{-i\omega_n(k-j)t}  ~b^g_k(t).
	\end{align}
\end{widetext}
We will assume, based on the discussion in subsection (\ref{perturb}), that the effect of the $\hbar W(t)$ term can be properly accounted for by first-order perturbation. In the first order, the coefficients are:
\begin{eqnarray}\label{PertQN}
	b^{e}_0(t)^{(1)} = b^e_0(0)[1 + i\zeta_1(t)] -  ib^g_0(0) \zeta_2(t) 
	\\
	b^g_0(t)^{(1)} = b^g_0(0)[1 - i\zeta_1(t)] -  ib^e_0(0) \zeta^*_2(t) 
\end{eqnarray}
In the second order:
\begin{eqnarray}\label{PertQN2}
	b^{e}_0(t)^{(2)} = ib^e_0(0) \chi \frac{t}{2}
	\\
	b^g_0(t)^{(2)} = -i b^g_0(0) \chi \frac{t}{2}
\end{eqnarray}
\begin{widetext}
	\begin{eqnarray*}
		b^e_1(t)^{(2)}
		= -i\frac{\Omega_n t}{2}e^{-i\phi_0} 
		~\mathcal{F}\!\left[ b^g_0(0)[1 - i\zeta_1] -  ib^e_0(0) \zeta^*_2 \right]\!
		\big(\!-\omega_n \big)   
		\\ +
		i\frac{\Omega_n \cos\theta}{2}t e^{-i\phi_0} 
		~\mathcal{F}\!\left[ b^g_0(0)[1 - i\zeta_1] -  ib^e_0(0) \zeta^*_2 \right]\!
		\big(\!-\omega_n \big)   
		+
		i\frac{\Omega_n \cos\theta}{2}t e^{i\phi_0} 
		~\mathcal{F}\!\left[ b^g_0(0)[1 - i\zeta_1] -  ib^e_0(0) \zeta^*_2 \right]\!
		\big(\!-\omega_n + 2\omega \big)   
		\\ -
		i\frac{\Omega_n \sin(2\theta)}{4}t e^{i\phi_0} 
		~\mathcal{F}\!\left[ b^e_0(0)[1 + i\zeta_1] -  ib^g_0(0) \zeta_2 \right]\!
		\big(\!-\omega_n -\omega \big)   
		-
		i\frac{\Omega_n \sin(2\theta)}{4}t e^{-i\phi_0} 
		~\mathcal{F}\!\left[ b^e_0(0)[1 + i\zeta_1] -  ib^g_0(0) \zeta_2 \right]\!
		\big(\!-\omega_n +\omega \big)   
	\end{eqnarray*}
	\begin{eqnarray*}
		b^g_1(t)^{(2)} 
		= -i\frac{\Omega_n t}{2}e^{-i\phi_0} 
		~\mathcal{F}\!\left[  b^e_0(0)[1 + i\zeta_1] -  ib^g_0(0) \zeta_2 \right]\!
		\big(\!-\omega_n \big)   
		\\ +
		i\frac{\Omega_n \cos\theta}{2}t e^{-i\phi_0} 
		~\mathcal{F}\!\left[  b^e_0(0)[1 + i\zeta_1] -  ib^g_0(0) \zeta_2 \right]\!
		\big(\!-\omega_n \big)   
		+
		i\frac{\Omega_n \cos\theta}{2}t e^{i\phi_0} 
		~\mathcal{F}\!\left[  b^e_0(0)[1 + i\zeta_1] -  ib^g_0(0) \zeta_2 \right]\!
		\big(\!-\omega_n + 2\omega \big)   
		\\ +
		i\frac{\Omega_n \sin(2\theta)}{4}t e^{i\phi_0} 
		~\mathcal{F}\!\left[ b^g_0(0)[1 - i\zeta_1] -  ib^e_0(0) \zeta^*_2 \right]\!
		\big(\!-\omega_n -\omega \big)   
		+
		i\frac{\Omega_n \sin(2\theta)}{4}t e^{-i\phi_0} 
		~\mathcal{F}\!\left[ b^g_0(0)[1 - i\zeta_1] -  ib^e_0(0) \zeta^*_2 \right]\!
		\big(\!-\omega_n +\omega \big)   
	\end{eqnarray*}
\end{widetext}
From further investigation, it is easy to check that up to the second power of time, the coefficients can be approximated as 
\begin{eqnarray}
	b_1^{e}(t)\approx b_1^{e}(t)^{(2)} 
	\big[1 + i \chi t\big] ,
	\\
	b_1^{g}(t)\approx b_1^{g}(t)^{(2)} 
	\big[1 - i \chi t\big] .
\end{eqnarray} 
We can infer that for the Raman-lines
\begin{equation*}
 b_1^{(e/g)}(t) \sim (1-e^{\pm i \chi t}) + o(\Omega^2)
\end{equation*} 
   approximately holds, that is, for interaction times the multiple of $2\pi/\chi$, the $b_1^{(e/g)}(t)$ coefficient is zero. A plot the value of $1/\chi$ as a function of the amplitude under resonant excitation can be seen on Fig.(\ref{fig:period}). 

%The same does not neccesserily stands for $b_2^{(e/g)}(t)$ and other coefficients of higher photon-number states. 

\section{Functions $\zeta_1$, $\zeta_2$, and $\chi$}\label{perturb}

Following the same notations as in the Appendix of \cite{GFV21Q}, we can analyse the semi-analytic model. The dynamical equations, in the special case of $\delta\omega=0$ become:

\begin{widetext}
	\begin{eqnarray*}\label{analdynequ3} 
		i \dot{b}^e(t) = b^e(t) \omega_0 \sum^{\infty}_{n=1}\bigg[ 
		J_{2n}(\eta) \cos(2\theta) \cos[2n(\omega t + \phi_0)]
		+ \frac{J_{2n+1}(\eta)}{2} \sin(2\theta) \big( \cos[2n(\omega t + \phi_0)] 
		- \cos[(2n+2)(\omega t + \phi_0)] \big)
		\bigg]
		\nonumber\\ 
		+  b^g(t) \omega_0 e^{i\omega t}
		\sum_{n=1}^{\infty} \bigg[ 
		J_{2n}(\eta) \sin(2\theta) \cos[2n(\omega t + \phi_0 )]
		- \frac{J_{2n+1}(\eta)}{2} \cos(2\theta) \big( \cos[2n (\omega t + \phi_0)] 
		- \cos[(2n+2)(\omega t + \phi_0)] \big)
		\nonumber\\ 
		+ i\frac{J_{2n+1}(\eta)}{2} \big( \sin[(2n+2)(\omega t + \phi_0 )] + \sin[2n(\omega t + \phi_0)]  \big) \bigg]
	\end{eqnarray*}
	\begin{eqnarray*}\label{analdynequ4} 
		i \dot{b}^g(t) = -b^g(t) \omega_0 \sum^{\infty}_{n=1}\bigg[ 
		J_{2n}(\eta) \cos(2\theta) \cos[2n(\omega t + \phi_0)]
		+ \frac{J_{2n+1}(\eta)}{2} \sin(2\theta) \big( \cos[2n(\omega t + \phi_0)] 
		- \cos[(2n+2)(\omega t + \phi_0)] \big)
		\bigg]
		\nonumber\\ 
		+  b^e(t) \omega_0 e^{-i\omega t}
		\sum_{n=1}^{\infty} \bigg[ 
		J_{2n}(\eta) \sin(2\theta) \cos[2n(\omega t + \phi_0)]
		- \frac{J_{2n+1}(\eta)}{2} \cos(2\theta) \big( \cos[2n(\omega t + \phi_0)] 
		- \cos[(2n+2)(\omega t + \phi_0)] \big)
		\nonumber\\ 
		- i\frac{J_{2n+1}(\eta)}{2} \big( \sin[(2n+2)(\omega t + \phi_0)] 
		+ \sin[2n(\omega t + \phi_0)]  \big) \bigg]
	\end{eqnarray*}
This special case corresponds to the parameters investigated in this paper.
Here we give the analytic expression of the first-order perturbation calculation results:
\begin{eqnarray}
	b^e(t) \approx b^e(0) +i b^e(0) \zeta_1(t) - i b^g(0) \zeta_2(t)
	\nonumber \\
	b^g(t) \approx b^g(0) -i b^g(0) \zeta_1(t) - i b^e(0) \zeta^*_2(t)
\end{eqnarray}
where we define the $\zeta_1(t)$ and $\zeta_2(t)$ expressions as:
	\begin{eqnarray}
		\zeta_1(t) \!=\!
		\omega_0  \sum^{\infty}_{n=1} \! \bigg[ 
		-\bigg( \frac{J_{2n}(\eta)}{2n\omega} \cos(2\theta)
		+ \frac{J_{2n+1}(\eta)}{4n\omega} \sin(2\theta) \bigg)
		\big( \sin[2n (\omega t + \phi_0) ] - \sin[2n \phi_0 ] \big)
		\\+
		\frac{J_{2n+1}(\eta)}{(4n+4)\omega} \sin(2\theta)
		\big( \sin[(2n+2)(\omega t + \phi_0) ] - \sin[(2n+2)\phi_0 ] \big)
		\bigg]
	\end{eqnarray}
	\begin{eqnarray}\hspace{-2.0cm}
		\zeta_2(t) \!=\!
		\omega_0  \sum^{\infty}_{n=1} \! 
		\left[ 
		\frac{i \left( J_{2n}(\eta)\sin(2\theta) - \frac{J_{2n+1}(\eta)}{2}\cos(2\theta) - J_{2n+1}(\eta) n \right) \left[ e^{i\omega t}\cos[2 n (\omega t + \phi)] - \cos[2 n \phi] \right] }
		{(4 n^2-1) \omega} \right.
		\\
		\left. 
		- \frac{
			\bigg( \frac{J_{2n+1}(\eta)}{2} + J_{2n+1}(\eta)\cos(2\theta)n 
			- 2 J_{2n}(\eta) \sin(2\theta)n \bigg) 
			\left[	e^{i \omega t} \sin[2 n (\omega t + \phi)] - \sin[2 n \phi]  \right]  }
		{(4 n^2-1) \omega}
		\right.
		\\
		+ \frac{i 
			\bigg( \frac{J_{2n+1}(\eta)}{2}\cos(2\theta) - J_{2n+1}(1 + n) \bigg)
			\left[e^{i\omega t} \cos[(2n+2) (\omega t+\phi)] - \cos[(2n+2) \phi] \right]}
		{(1 + 2 n) (3 + 2 n) \omega} 
		\\
		\left. - \frac{ 
			\bigg( \frac{J_{2n+1}(\eta)}{2} - J_{2n+1}\cos(2\theta)(1 + n) \bigg)
			\big[ e^{i\omega t} \sin[(2n + 2) (\omega t + \phi )] - \sin[(2n + 2) \phi] \big]
		}{(1 + 2 n) (3 + 2 n) \omega} \right]
	\end{eqnarray}
Our goal at this point is to express the timescale associated with the "period" of the dynamics of the squeezing. For this, we estimate the functional dependence of $\chi$. At first step, we do this by leaving aside most of the non-resonant terms, and get:
	\begin{eqnarray*}
		\frac{\dot{b}^e_0(t)}{b^e_0(0)} \approx \omega^2_{0} 
		\left\{
		\frac{-i \left( J_{2}(\eta)\sin(2\theta) - \frac{J_{3}(\eta)}{2}\cos(2\theta) - J_{3}(\eta) \right) }
		{3 \omega}  \cos[2 (\omega t + \phi)] \right.
		\\  \left. -
		i \sum_{n=1}^{\infty} \left[ 
		\frac{ \left( J_{2n+2}(\eta)\sin(2\theta) - J_{2n+3}(\eta)\bigg[\frac{\cos(2\theta)}{2}+(n+1)\bigg] \right)  }
		{(4 (n+1)^2-1) \omega} 
		+
		\frac{
			\bigg( \frac{J_{2n+1}(\eta)}{2}\cos(2\theta) - J_{2n+1}(1 + n) \bigg)}
		{(1 + 2 n) (3 + 2 n) \omega} \right]
		 \right.
\\
\times  \cos[(2n+2) (\omega t+\phi)] 
		\left. 
		+ \frac{
			\bigg( \frac{J_{3}(\eta)}{2} + J_{3}(\eta)\cos(2\theta) 
			- 2 J_{2}(\eta) \sin(2\theta) \bigg)  }
		{3 \omega} \sin[2 (\omega t + \phi)]
		\right.
\end{eqnarray*}
\begin{eqnarray*}		
		\\+  \sum_{n=1}^{\infty}  \left[ \frac{
			\bigg( J_{2n+3}(\eta)\bigg(\frac{1}{2} + \cos(2\theta)(n+1) \bigg)
			- 2 J_{2n+2}(\eta) \sin(2\theta)(n+1) \bigg)  }
		{(4 (n+1)^2-1) \omega}
		\left.   + \frac{ 
			\bigg( \frac{J_{2n+1}(\eta)}{2} - J_{2n+1}\cos(2\theta)(1 + n) \bigg) 
		}{(1 + 2 n) (3 + 2 n) \omega} \right] \right.
		\\ \left.
		\times \sin[(2n + 2) (\omega t + \phi )]
		\right\}
		\\
		\times \left\{
		\bigg(  J_{2}(\eta) \sin(2\theta)
		- \frac{J_{3}(\eta)}{2} \cos(2\theta) \bigg)
		\cos[2(\omega t + \phi_0 )] \right.
		\\
		+ \sum_{n=1}^{\infty} \left[ 
		J_{2n+2}(\eta) \sin(2\theta)
		- \frac{J_{2n+3}(\eta)}{2} \cos(2\theta) 
		-  \frac{J_{2n+1}(\eta)}{2} \cos(2\theta) \right]
		\cos[(2n+2)(\omega t + \phi_0)] 
		\\ 
		\left. + 
		i\frac{J_{3}(\eta)}{2}\sin[2(\omega t + \phi_0 )] 
		+ i\sum_{n=1}^{\infty} \bigg( \frac{J_{2n+1}(\eta)}{2} + \frac{J_{2n+3}(\eta)}{2} \bigg) \sin[(2n+2)(\omega t + \phi_0 )]  \bigg]
		\right\}
	\end{eqnarray*}
from which the resonant terms are simple to identify.
	\begin{eqnarray*}
		\frac{b^e_0(t)}{b^e_0(0)}\approx \omega^2_{0} \frac{t}{2}
		\left\{
		\frac{-i \left( J_{2}(\eta)\sin(2\theta) - \frac{J_{3}(\eta)}{2}\cos(2\theta) - J_{3}(\eta) \right) }
		{3 \omega} 
		\bigg(  J_{2}(\eta) \sin(2\theta)
		- \frac{J_{3}(\eta)}{2} \cos(2\theta) \bigg) \right.
		\\  \left. 
		-i \sum_{n=1}^{\infty} \left[ 
		\frac{ \left( J_{2n+2}(\eta)\sin(2\theta) - J_{2n+3}(\eta)\bigg[\frac{\cos(2\theta)}{2}+(n+1)\bigg] \right)  }
		{(4 (n+1)^2-1) \omega} 
		+
		\frac{
			\bigg( \frac{J_{2n+1}(\eta)}{2}\cos(2\theta) - J_{2n+1}(1 + n) \bigg)}
		{(1 + 2 n) (3 + 2 n) \omega} \right]
		\right.
		\\ \times \left[ 
		J_{2n+2}(\eta) \sin(2\theta)
		- \frac{J_{2n+3}(\eta)}{2} \cos(2\theta) 
		-  \frac{J_{2n+1}(\eta)}{2} \cos(2\theta) \right]
		\left. 
		+ i\frac{
			\bigg( \frac{J_{3}(\eta)}{2} + J_{3}(\eta)\cos(2\theta) 
			- 2 J_{2}(\eta) \sin(2\theta) \bigg)  }
		{3 \omega} 	\frac{J_{3}(\eta)}{2}
		\right.
		\\+ i \sum_{n=1}^{\infty}  \left[ \frac{
			\bigg( J_{2n+3}(\eta)\bigg(\frac{1}{2} + \cos(2\theta)(n+1) \bigg)
			- 2 J_{2n+2}(\eta) \sin(2\theta)(n+1) \bigg)  }
		{(4 (n+1)^2-1) \omega}
		\left.   + \frac{ 
			\bigg( \frac{J_{2n+1}(\eta)}{2} - J_{2n+1}\cos(2\theta)(1 + n) \bigg) 	}{(1 + 2 n) (3 + 2 n) \omega} \right] \right.
		\\ 
		\times	\left. 
		\bigg( \frac{J_{2n+1}(\eta)}{2} + \frac{J_{2n+3}(\eta)}{2} \bigg)
		\right\}
	\end{eqnarray*}
	
	\begin{eqnarray*}
		\chi = \omega^2_{0} 
		\left\{
		\frac{ -\left( J_{2}(\eta)\sin(2\theta) - \frac{J_{3}(\eta)}{2}\cos(2\theta) - J_{3}(\eta) \right) }
		{3 \omega} 
		\bigg(  J_{2}(\eta) \sin(2\theta)
		- \frac{J_{3}(\eta)}{2} \cos(2\theta) \bigg) \right.
		\\  \left. -
		\sum_{n=1}^{\infty} \left[ 
		\frac{ \left( J_{2n+2}(\eta)\sin(2\theta) - J_{2n+3}(\eta)\bigg[\frac{\cos(2\theta)}{2}+(n+1)\bigg] \right)  }
		{(4 (n+1)^2-1) \omega} 
		+
		\frac{
			\bigg( \frac{J_{2n+1}(\eta)}{2}\cos(2\theta) - J_{2n+1}(1 + n) \bigg)}
		{(1 + 2 n) (3 + 2 n) \omega} \right]
		\right.
		\\ \times \left[ 
		J_{2n+2}(\eta) \sin(2\theta)
		- \frac{J_{2n+3}(\eta)}{2} \cos(2\theta) 
		-  \frac{J_{2n+1}(\eta)}{2} \cos(2\theta) \right]
		\left. 
		+ \frac{
			\bigg( \frac{J_{3}(\eta)}{2} + J_{3}(\eta)\cos(2\theta) 
			- 2 J_{2}(\eta) \sin(2\theta) \bigg)  }
		{3 \omega} 	\frac{J_{3}(\eta)}{2}
		\right.
		\\+  \sum_{n=1}^{\infty}  \left[ \frac{
			\bigg( J_{2n+3}(\eta)\bigg(\frac{1}{2} + \cos(2\theta)(n+1) \bigg)
			- 2 J_{2n+2}(\eta) \sin(2\theta)(n+1) \bigg)  }
		{(4 (n+1)^2-1) \omega}
		\left.   + \frac{ 
			\bigg( \frac{J_{2n+1}(\eta)}{2} - J_{2n+1}\cos(2\theta)(1 + n) \bigg) 	}{(1 + 2 n) (3 + 2 n) \omega} \right] \right.
		\\ 
		\times	\left. 
		\bigg( \frac{J_{2n+1}(\eta)}{2} + \frac{J_{2n+3}(\eta)}{2} \bigg)
		\right\}
	\end{eqnarray*}
\end{widetext}

\end{document}